\documentclass{ws-mplb}

\usepackage{amsmath,amssymb}
\usepackage{graphicx}
\usepackage{dcolumn}
\usepackage{bm}
\usepackage[dvips]{color}

\def\rnum#1{\expandafter{%
\romannumeral #1}}
\def\Rnum#1{\uppercase\expandafter{%
\romannumeral #1}}
\newcommand{\bol}[1]{\boldsymbol #1}

\begin{document}

\markboth{Masahiro Sato}
{Phase Diagram}

%
\catchline{}{}{}{}{}
%

\title{Competing Phases in Spin-$\frac{1}{2}$ $J_1$-$J_2$ Chain with 
Easy-Plane Anisotropy}

\author{Masahiro Sato$^1$, Shunsuke Furukawa$^2$, Shigeki Onoda$^3$ 
and Akira Furusaki$^3$}
\address{${}^1$Department of Physics and Mathematics, Aoyama Gakuin University, 
Sagamihara, Kanagawa 252-5258, Japan\\
${}^2$Department of Physics, University of Toronto, Toronto, Ontario, 
Canada M5S 1A7\\
${}^3$Condensed Matter Theory Laboratory, RIKEN, Wako, Saitama 351-0198, Japan}

\maketitle

\begin{history}
\received{(Day Month Year)}
\revised{(Day Month Year)}
\end{history}

\begin{abstract}
We summarize our theoretical findings on the ground-state phase diagram 
of the spin-$\frac12$ XXZ chain having competing
nearest-neighbor ($J_1$) and antiferromagnetic next-nearest-neighbor ($J_2$)
couplings. 
Our study is mainly concerned with the case of ferromagnetic $J_1$,
and the case of 
antiferromagnetic $J_1$ is briefly reviewed for comparison. 
The phase diagram contains a rich variety of phases 
in the plane of $J_1/J_2$ versus the XXZ anisotropy $\Delta$:
vector-chiral phases, N\'eel phases, several dimer phases, and 
Tomonaga-Luttinger liquid phases. 
We discuss the vector-chiral order that appears for
a remarkably wide parameter space,
successive N\'eel-dimer phase transitions,
and an emergent nonlocal string order in a narrow region of 
ferromagnetic $J_1$ side. 
\end{abstract}


\section{Introduction}
\label{Intro}
Frustrated spin systems have been a subject of interest
because of their rich physics arising from
competing interactions and quantum/thermal fluctuations.\cite{diep,balents}
One-dimensional (1D) frustrated spin models 
provide one of the prototypical families to theoretically 
study unconventional orders with high accuracy. 
We here focus on the spin-$\frac12$ XXZ chain 
with nearest-neighbor (NN) exchange coupling $J_1$ and 
next-nearest-neighbor (NNN) coupling $J_2$. 
The Hamiltonian is given by 
\begin{eqnarray}
\cal H&=&\sum_{n=1,2}\sum_{j}J_n(S_j^x S_{j+n}^x+S_j^y S_{j+n}^y
+\Delta S_j^z S_{j+n}^z). 
\label{eq:Ham}
\end{eqnarray}
Here $\Delta$ is the XXZ anisotropy, and we will consider the 
easy-plane region $0\leq \Delta\leq 1$. 
In the case of antiferromagnetic (AF) $J_2>0$, 
NN and NNN couplings are geometrically frustrated 
irrespective of the sign of $J_1$. 
This model has been studied in detail when $J_1$ and $J_2$ are both AF, 
and its ground-state and low-energy properties are now
well understood.\cite{MG69,haldane82,NO94,white96,nersesyan98,hikihara01}
By contrast, relatively less is understood for
the case of ferromagnetic (FM) $J_1$ and AF $J_2$
despite earlier studies.\cite{tonegawa90,chubukov91,itoi,SA01}
Recently, interest has been growing in this FM $J_1$ case
because of its possible relevance to several quasi-1D edge-sharing cuprates 
[$\rm LiCu_2O_2$ (Ref.~\cite{masuda05}), 
$\rm LiCuVO_4$ (Ref.~\cite{enderle05}), 
$\rm Rb_2Cu_2Mo_3O_{12}$ (Ref.~\cite{hase04}), 
$\rm PbCuSO_4(OH)_2$ (Ref.~\cite{kamieniarz,baran}), etc.].
Some of these compounds exhibit multiferroic behavior in 
low-temperature spiral spin ordered phases, 
where the long-range order (LRO) of the vector spin chirality 
produces the electric polarization.\cite{katsura,jia} 
Motivated by these developments,
we have recently performed intensive studies on 
the ground-state phase diagram of the spin-$\frac12$ $J_1$-$J_2$ chain
(\ref{eq:Ham}) with FM $J_1<0$ and AF $J_2>0$.\cite{FSSO08,FSO10,FSF10,FSO} 
In particular, among various phases which we have successfully 
characterized, the emergent LRO of the vector spin chirality for
ferromagnetic $J_1$ with a weak easy-plane anisotropy would be relevant
to quasi-1D multiferroic cuprates,~\cite{FSO10} when interchain
couplings are taken into account.

The effect of external magnetic field is another interesting direction of
research for the frustrated $J_1$-$J_2$ spin chain systems. Recently, 
field-induced Tomonaga-Luttinger liquid (TLL) phases 
with spin multipolar quasi LRO have been theoretically 
investigated in the model (\ref{eq:Ham}),\cite{hikihara08,sudan09} 
and it has been predicted that these multipolar phases show characteristic 
dynamical spin response which can be observed in NMR and neutron-scattering 
experiments.\cite{sato09,sato10,sato} 
Throughout this paper, however, we restrict ourselves to the 
case at zero magnetic field. 
In the following sections, we review 
our recent results\cite{FSSO08,FSO10,FSF10,FSO} on
the phase diagram and 
the characteristic features of the phases in the model (\ref{eq:Ham}).
Including the parameter space of AF $J_1>0$, the phase diagram 
consists of at least six (and presumably more) kinds of distinct phases.

\section{Competition between chiral and dimer orders}
\label{sec:intermediate}
In this section, we consider the regime $-4< J_1/J_2< 4$ with $J_2>0$, 
where the classical ground state has an incommensurate spin spiral structure.
This spiral state has a non-vanishing vector spin chirality 
$\kappa^z_j=\langle({\bol S}_j\times{\bol S}_{j+1})^z\rangle\neq 0$. 
In the quantum case, a true spiral LRO with broken spin
rotational symmetry is difficult to occur in 1+1 dimensions,\cite{Momoi}
but a vector chiral LRO is allowed for $\Delta\ne 1$ since 
it only breaks a discrete $\mathbb{Z}_2$ symmetry.\cite{nersesyan98} 
As explained below, 
this vector chiral order competes with several kinds of dimer orders 
driven by quantum fluctuations.

\begin{figure}[t]
\centerline{\psfig{file=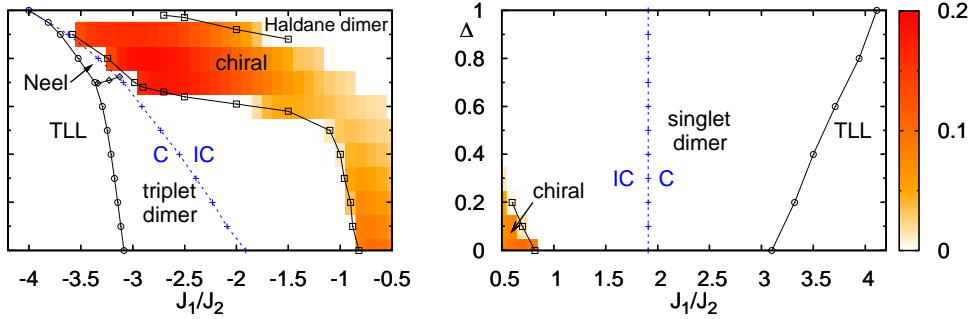,width=13cm}}
\caption{Ground-state phase diagram of the $J_1$-$J_2$ 
model~(\ref{eq:Ham}) in the classically spiral regime 
$-4\lesssim J_1/J_2 \lesssim4$ with $J_2>0$. 
Strength of the vector chirality 
$\langle({\bol S}_j\times{\bol S}_{j+1})^z\rangle$ calculated by iTEBD is also 
plotted with red color. 
The small N\'eel phase in the FM-$J_1$ side will be discussed 
in Sec.~\ref{sec:NeelDimer}.}
\label{fig:phase1}
\end{figure}

The ground-state phase diagram 
is presented in Fig.~\ref{fig:phase1}. 
The AF-$J_1$ side has been 
already established by several theoretical works. 
A strong enough AF NNN interaction $J_2$ causes 
a Kosteriz-Thouless (KT) transition from the TLL phase (connected to
a single XXZ spin chain with the NN exchange coupling $J_1$)
to a dimerized phase 
with spontaneously broken translational symmetry,
in which the ground state is doubly degenerate.\cite{haldane82,NO94} 
This dimer phase occupies a large part of the classical spiral 
regime $J_1/J_2<4$ of the AF-$J_1$ side. 
In fact, the ground state on the line $J_1/J_2=2$ and with $\Delta>-1/2$ 
(extension of the Majumder-Ghosh model\cite{MG69} at $\Delta=1$) 
is given by a product state of singlet bonds, 
$|GS\rangle=\prod_{j=\rm even}
({|\!\!\uparrow\rangle_j|\!\!\downarrow\rangle_{j+1}}
-{|\!\!\downarrow\rangle_j|\!\!\uparrow\rangle_{j+1}})$ or 
$\prod_{j=\rm odd}
(|\!\!\uparrow\rangle_j|\!\!\downarrow\rangle_{j+1}
-|\!\!\downarrow\rangle_j|\!\!\uparrow\rangle_{j+1})$, where 
$|\!\!\uparrow\rangle_j~(|\!\!\downarrow\rangle_j)$ is the eigenstate of 
$S_j^z$ with the eigenvalue $+1/2$ $(-1/2)$.  
For later convenience, we introduce the $xy$ and $z$ 
components of dimer order parameters
\begin{eqnarray}
D^{xy}_j &=& \langle(S_{j-1}^xS_j^x+S_{j-1}^yS_j^y)
-(S_j^xS_{j+1}^x+S_j^yS_{j+1}^y)\rangle,\\
D_j^z &=& \langle S_{j-1}^zS_j^z-S_j^zS_{j+1}^z\rangle.
\end{eqnarray}
For the above product state, we can easily show that 
(i) ${\rm sign}(D^{xy}_j)={\rm sign}(D^{z}_j)$, and (ii) 
the energy density $\langle {\bol S}_j\cdot{\bol S}_{j+1} \rangle$ 
on the dimerized bond is negative.
These properties persist
in the whole region of the dimer phase in the AF-$J_1$ side. 
We call this phase the ``singlet-dimer'' phase. 
In this phase, there is a Lifshitz line across which
the short-range spin correlation changes its character 
from commensurate to incommensurate 
(C and IC in Fig.~\ref{fig:phase1}).\cite{note1}

In the weak $J_1$ region where $0<J_1/J_2\lesssim0.8$, 
a vector-chiral phase appears\cite{nersesyan98,hikihara01}
in which the $\mathbb{Z}_2$ parity symmetry is spontaneously broken,
and a nonvanishing and spatially uniform average of 
$\kappa^z_j=\langle({\bol S}_j\times{\bol S}_{j+1})^z\rangle$ is found.
In contrast to the dimer phase, the vector-chiral phase has a gapless
excitation mode, 
and both the longitudinal and transverse spin correlation functions 
decay in a power-law fashion. In particular, the transverse spin 
correlator has an incommensurate oscillating factor.

Now, we turn to the FM-$J_1$ side, which has been investigated in our 
recent works.\cite{FSO10,FSO}  
Phase boundaries are numerically determined by using infinite time 
evolving block decimation (iTEBD) method\cite{vidal} and 
numerical diagonalization. 
The phase diagram in Fig.~\ref{fig:phase1} clearly shows that
for $J_1<0$ the vector-chiral phase 
with $\kappa^z_j\neq 0$ appears for much broader parameter space
than in the AF-$J_1$ case ($J_1>0$),
and extends up to the vicinity of the SU(2) line $\Delta=1$ 
for moderate values of $|J_1|/J_2$. 
This result naturally explains why quasi-1D $J_1$-$J_2$ magnets 
with FM $J_1$ coupling often show a spiral spin order at low
temperatures,\cite{masuda05,enderle05} 
while with AF $J_1$ coupling no quasi-1D magnet 
with a spiral order is found so far. 
Namely, a chiral ordered state appearing for realistically small
easy-plane anisotropy $1-\Delta\ll 1$ in the FM-$J_1$ side
can be easily promoted to a 3D spiral ordered state by the addition of
weak interchain couplings, 
while a gapped dimer state in the AF-$J_1$ side will be robust
against weak 3D couplings.

On the FM-$J_1$ side, there appear two distinct types of dimer phases
between which the vector-chiral phase intervene in Fig.~\ref{fig:phase1}. 
The wider dimer phase appearing for strong easy-plane anisotropy 
$0\leq \Delta \lesssim 0.6$ 
can be easily understood as follows.\cite{chubukov91} 
In the XY limit $\Delta=0$, the spin chain with an AF NN coupling $J_1=J>0$
can be mapped to the same spin chain with the opposite sign of NN coupling
($J_1=-J<0$) through $\pi$ rotation of spins
around $S^z$ axis on every second site.
By this transformation a singlet dimer
$|\!\!\uparrow\downarrow\rangle-|\!\!\downarrow\uparrow\rangle$ 
on a bond is changed into a triplet state
$|\!\!\uparrow\downarrow\rangle+|\!\!\downarrow\uparrow\rangle$. Therefore, 
the ground state at $(J_1/J_2,\Delta)=(-2,0)$ is a product state of 
triplet bonds, $|GS\rangle=\prod_{j=\rm even}
(|\!\!\uparrow\rangle_j|\!\!\downarrow\rangle_{j+1}
+|\!\!\downarrow\rangle_j|\!\!\uparrow\rangle_{j+1})$ or 
$\prod_{j=\rm odd}
({|\!\!\uparrow\rangle_j|\!\!\downarrow\rangle_{j+1}}
+|\!\!\downarrow\rangle_j|\!\!\uparrow\rangle_{j+1})$. 
These states show ${\rm sign}(D^{xy}_j)=-{\rm sign}(D^{z}_j)$, 
and this property persists in the whole region of this dimer phase. 
We therefore call this phase the ``triplet-dimer'' phase.  
The singlet- and triplet-dimer phases can be distinguished 
by the signs of the dimer order parameters.

\begin{figure}[bth]
\psfig{file=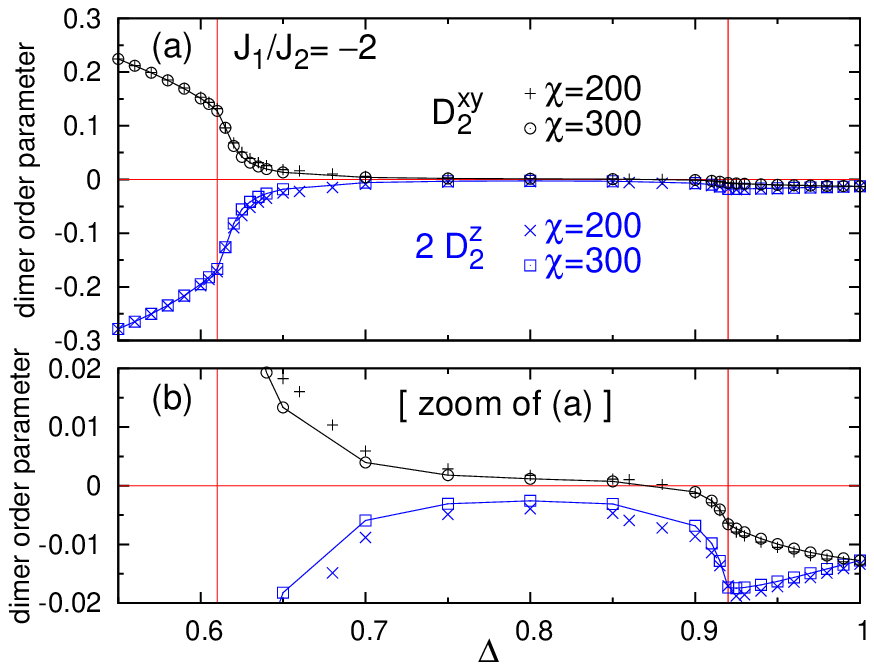,width=6cm}\hspace{0.5cm}
\psfig{file=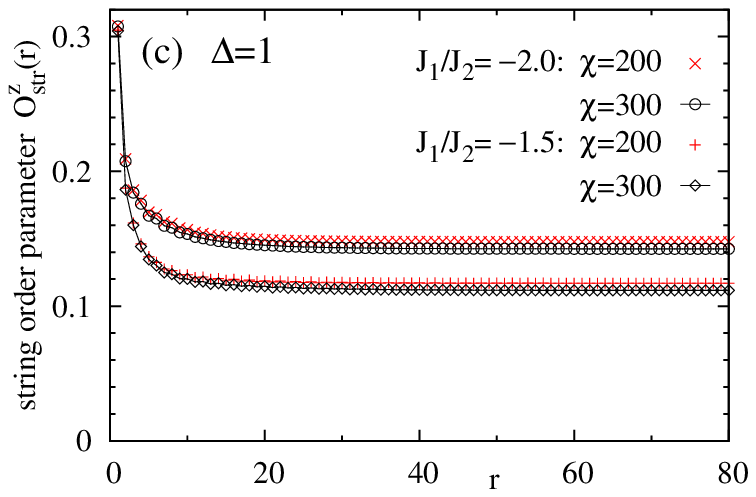,width=6cm}
\caption{(a) $\Delta$ dependence of dimer order parameters $D^{xy}_j$ 
and $D^z_j$ on $J_1/J_2=-2$ line, calculated by iTEBD method with 
Schmidt rank $\chi=200$ and 300. 
The chirality $\kappa^z_j$ is finite between 
two vertical lines at $\Delta\simeq 0.61$ and $\Delta\simeq 0.92$. 
Panel (b) is a zoom of panel (a). 
(c) String order parameter~(\ref{eq:string}) at points 
$(J_1/J_2,\Delta)=(-2,1)$ and $(-1.5,1)$ of the dimer phase, 
calculated by iTEBD method.}
\label{fig:dimer_string}
\end{figure}

The nature of the other dimer phase around the SU(2) line $\Delta=1$ has
long been controversial. 
We have clarified some 
characteristic properties of this phase by applying unbiased iTEBD method. 
As shown in Fig.~\ref{fig:dimer_string}(a)(b) where dimer correlations
are plotted for one of doubly degenerate ground states,
the magnitude of dimer orders 
is quite tiny and ${\rm sign}(D^{xy}_j)={\rm sign}(D^{z}_j)$ is realized. 
The latter property is the same as in the singlet-dimer phase. 
However, we have found that
$\langle {\bol S}_{j}\cdot{\bol S}_{j+1} \rangle$ on a dimerized bond is
positive, while it is negative in the singlet-dimer phase. 
This FM dimerization suggests 
that an effective spin-1 degree of freedom emerges on each of dimerized bonds. 
Therefore, we can expect a realization of a valence-bond-solid (VBS) 
state\cite{AKLT} like in a spin-1 AF chain.
In fact, the emergence of an effective spin-1 chain is very natural 
for FM $J_1$ coupling.
To judge whether the dimer phase around $\Delta=1$ can be well approximated 
by a VBS state, we calculate the string order parameter\cite{NR89,KT92}
\begin{eqnarray}
\label{eq:string}
O_{\rm str}^\alpha(j-k)&=&-\left\langle (S_{2j}^\alpha+S_{2j+1}^\alpha)
\exp\left[i\pi\sum_{l=j+1}^{k-1}(S_{2l}^\alpha+S_{2l+1}^\alpha)\right]
(S_{2k}^\alpha+S_{2k+1}^\alpha)\right\rangle,
\end{eqnarray}
where we have assumed that bonds $(2l,2l+1)$ are dimerized. 
Figure~\ref{fig:dimer_string}(c) clearly shows that the string 
parameter is long-range ordered.
By contrast, the string order parameter defined on the non-dimerized bonds
$(2l-1,2l)$ is found to be short-ranged.
We may expect that the dimer phase around $\Delta=1$ should be adiabatically 
connected to a spin-1 AF chain if we introduce a strong FM bond 
alternation on $J_1$ bonds. 
(Note, however, that the dimerization is a spontaneous symmetry 
breaking in our model.)
We thus call this phase the ``Haldane-dimer'' phase. Here, we leave the
issue of whether this Haldane-dimer phase is adiabatically connected to
the singlet-dimer phase or not for future studies.\cite{FSO}

Figure \ref{fig:dimer_string} suggests 
the presence of another type of dimer phase,
in a narrow region of parameter
space ($0.61\lesssim\Delta\lesssim 0.65$ at $J_1/J_2=-2$),
which is characterized by
coexisting dimer order $D^{xy}_j,D^z_j$
and vector chirality $\kappa^z_j$.
Let us call this phase the ``chiral-dimer'' phase. 
We note that evaluated dimer order parameters in
Fig.~\ref{fig:dimer_string} cannot be used to determine the phase
boundary between the chiral-dimer and vector-chiral phases.
Further calculations are ongoing to verify the existence and to
determine the range of this phase. 
More detailed discussions on the chiral, Haldane-dimer and 
chiral-dimer phases will be given in Ref.~\cite{FSO}.

\section{Strong $|\bol J_{\bol 1}|$ region: successive N\'eel-dimer transitions}
\label{sec:NeelDimer}
\begin{figure}[th]
\psfig{file=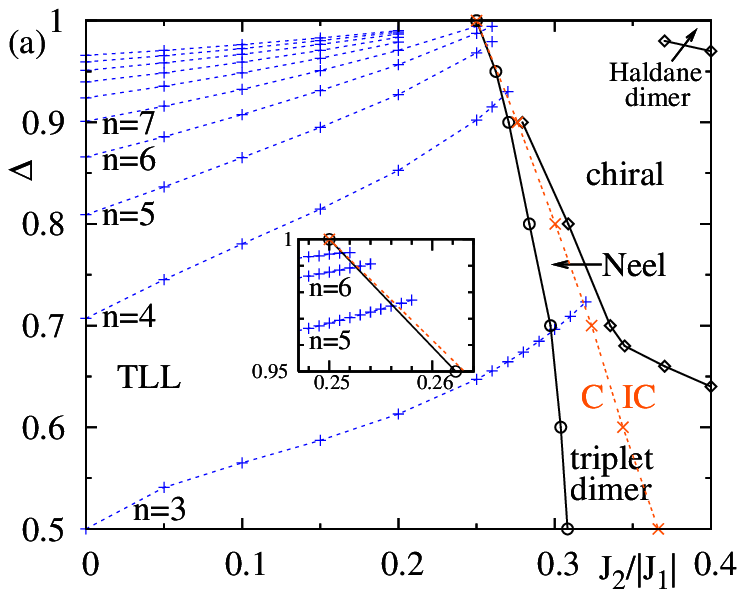,width=5.5cm}\hspace{0.3cm}
\psfig{file=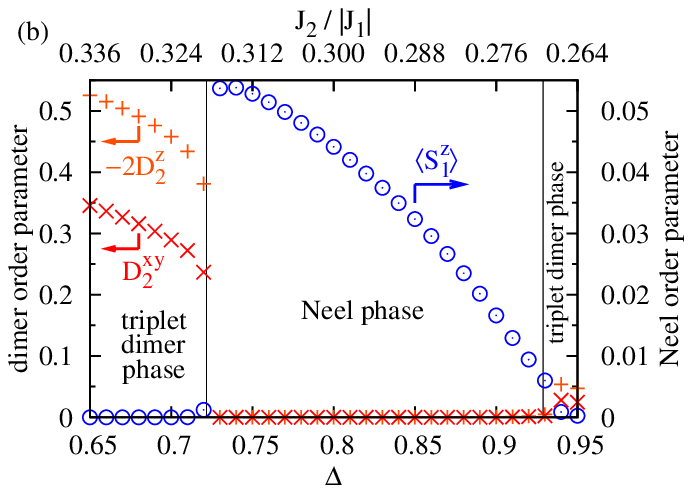,width=6.5cm}
\caption{(a) Ground-state phase diagram of the model~(\ref{eq:Ham}) 
with FM $J_1<0$ for relatively small $J_2/|J_1|$. 
(b) Dimer and N\'eel order parameters, calculated by iTEBD method, 
along the Lifshitz line. 
The vertical lines indicate the phase boundaries determined 
by the level spectroscopy.}
\label{fig:Neel_dimer}
\end{figure}

In this section, we focus on the narrow region between the chiral and TLL phase 
around $-4\lesssim J_1/J_2\lesssim-3$. 
For such large negative $J_1/J_2$, 
a single $J_1$ chain with $J_2=0$, which is exactly solvable, 
becomes a useful starting point. 
The low-energy effective Hamiltonian for the $J_1$ chain is 
a free boson (i.e., TLL) model with a nonlinear (vertex) term:
\begin{eqnarray}
\label{eq:boson}
{\cal H}_{\rm eff}&=&
 \frac{v}2 \left[
 K \left(\partial_x\theta\right)^2
  + \frac1K \left(\partial_x\phi\right)^2
 \right]
 - \frac{v\lambda}{2\pi} \cos(\sqrt{16\pi}\phi),
\end{eqnarray}
where $x=ja$ ($a$ is lattice spacing), 
$(\phi(x),\theta(x))$ is a pair of dual scalar fields, 
$K$ is the TLL parameter, $v$ is the spinon velocity, and 
$\lambda$ is the coupling constant of the perturbative vertex term. 
For the $J_1$ chain with $0\leq \Delta< 1$, 
the TLL parameter is given by 
$K=\pi/(2\cos^{-1}\Delta)$. 
Since $K>1$ in this case,  
the $\lambda$ term (scaling dimension $4K$) is irrelevant
in the renormalization-group sense, 
and a TLL phase is realized. Furthermore, 
the exact value of $\lambda$ is known\cite{lukyanov} and 
has an oscillating factor $-\sin(2\pi K)$ in the $J_1$ chain. 
Thus, $\lambda$ changes its sign and becomes zero when $2K=n$, namely, 
\begin{eqnarray}
\label{eq:zero}
\Delta=\cos(\pi/n), && n=3,4,\cdots. 
\end{eqnarray}
When small $J_2$ is introduced, the parameters $K$, $v$, and $\lambda$ 
generally change, and finally the $\lambda$ term becomes relevant 
and the TLL phase is destabilized towards gapped phases.\cite{haldane82,NO94} 
This boundary is determined in Ref.~\cite{SA01},  
and is plotted by circular symbols interpolated by solid lines in 
Fig.~\ref{fig:Neel_dimer}(a). 
For the effective theory~(\ref{eq:boson}) it is known that, 
if the $\lambda$ term is relevant, then positive $\lambda$ induces 
a N\'eel order with finite $\langle S_j^z\rangle=-\langle S_{j+1}^z\rangle$, 
while negative $\lambda$ induces 
a dimer order with finite $\langle {\bol S}_j\cdot{\bol S}_{j+1}
-{\bol S}_{j+1}\cdot{\bol S}_{j+2}\rangle$.  
Thus a point of $\lambda=0$ 
separates the dimer and N\'eel phases in the region with 
$4K<2$. Using the level spectroscopy method of Ref.~\cite{NO94}, 
we have determined the curves of $\lambda=0$,\cite{FSF10} 
which start from the points of 
Eq.~(\ref{eq:zero}) and are plotted as ``+'' symbols interpolated by broken lines. 
Remarkably, all the curves continue even outside the TLL phase. 
It means that successive dimer-N\'eel transitions 
occur as $\Delta$ is increased in the narrow region 
between TLL and chiral phases. In particular,
the emergence of N\'eel order along $S^z$ axis is nontrivial 
since it seems unfavored by both FM $J_1$ and AF $J_2$ 
couplings in the classical-spin picture.

The above argument has been based on the effective theory~(\ref{eq:boson}). 
Using unbiased iTEBD, we have also directly calculated the dimer and
N\'eel order parameters along the Lifshitz line, where the order
parameters are relatively large; see Fig.~\ref{fig:Neel_dimer}(b). 
We find that the transition points determined in
Fig.~\ref{fig:Neel_dimer} (a) and (b) are consistent.

\section{Summary}
\label{sec:sum}
In this paper, we have discussed the ground-state properties of 
a spin-$\frac{1}{2}$ $J_1$-$J_2$ chain (\ref{eq:Ham}) with easy-plane anisotropy, 
especially, for the FM-$J_1$ case. 
The ground-state phase diagram contains very rich physics: 
vector-chiral phase, four kinds of dimerized phases 
(singlet, triplet, Haldane and chiral dimers), N\'eel phases and TLL. 
Important findings in our recent studies on the FM-$J_1$ case include 
(i) remarkable stability of the vector-chiral phase even near 
$\Delta=1$,\cite{FSO10} 
(ii) a finite string order of the Haldane-dimer phase,\cite{FSO} 
and
(iii) successive N\'eel-dimer transitions.\cite{FSF10} 
Furthermore, our numerical result suggests 
the possible coexistence of chirality and dimerization 
in the narrow chiral-dimer phase.\cite{FSO}  
More detailed properties of vector-chiral, Haldane-dimer, and
chiral-dimer phases in FM-$J_1$ side will be discussed elsewhere.\cite{FSO}

\section*{Acknowledgments}
This work was supported by Grants-in-Aid for Scientific Research
from MEXT, Japan (Grants No.\ 21740295, and No.\ 22014016).



\begin{thebibliography}{0}
\bibitem{note1}
We have determined the C-IC boundary by looking at the peak
position of the in-plane spin structure factor.
\end{thebibliography}
\end{document}